\documentclass{PoS}

\title{High energy astrophysics with the next generation of radio
  astronomy facilities}

\ShortTitle{Next generation radio facilities}

\author{\speaker{Rob Fender}\\
        University of Southampton\\
        E-mail: \email{r.fender@soton.ac.uk}}

\abstract{High energy astrophysics has made good use of combined high
  energy (X-ray, $\gamma$-ray) and radio observations to uncover
  connections between outbursts, accretion, particle acceleration and
  kinetic feedback to the local ambient medium. In the field of
  microquasars the connections have been particularly
  important. However, radio astronomy has been relying on essentially
  the same facilities for the past $\sim 25$ years, whereas
  high-energy astrophysics, in particular space-based research, has
  had a series of newer and more powerful missions. In the next
  fifteen years this imbalance is set to be redressed, with a whole
  familiy of new radio facilities under development en route to the
  Square Kilometre Array (SKA) in the 2020s. In this brief review I
  will summarize these future prospects for radio astronomy, and focus
  on possibly the most exciting of the new facilities to be built in
  the next decade, the Low Frequency Array LOFAR, and its uses in high
  energy astrophysics.  }

\FullConference{VII Microquasar Workshop: Microquasars and Beyond\\
		 September 1-5 2008\\
		 Foca, Izmir, Turkey}

\begin{document}

\section{Introduction}

High-energy astrophysics in general, and work on X-ray binary jets
(`microquasars') in particular, has revealed important connections
between radio and X-ray emission in astronomical objects. X-rays trace
the instantaneous accretion rate, and radio emission the recent
history of matter ejection, thereby allowing a measure of how much of
the available accretion power is released in the form of radiation,
and how much in kinetically-dominated flows (or even advected across
an event horizon in the case of black holes).

In the early 1980s, when work on jets from X-ray binaries was first
beginning (e.g. Hjellming et al. 1980), the radio community were
commissioning a new and powerful facility, the Very Large Array (VLA).
The X-ray community were analysing the exciting results from {\em
  Einstein} and looking forward to the imminent launch of EXOSAT, each
of which pushed back the frontiers of X-ray astronomy. Nearly thirty
years later, in 2008, the X-ray community can look back on an exciting
three decades of new facilities, each (usually) more powerful than
those which came before. In the radio world, however, the most
powerful radio facility remains the VLA (very honourable mentions
should of course be given to newcomers such as ATCA and GMRT). Fig 1
compares the arrival of new facilities in radio and X-ray astronomy
over the past 40 years.

\begin{figure}
\includegraphics[width=15cm]{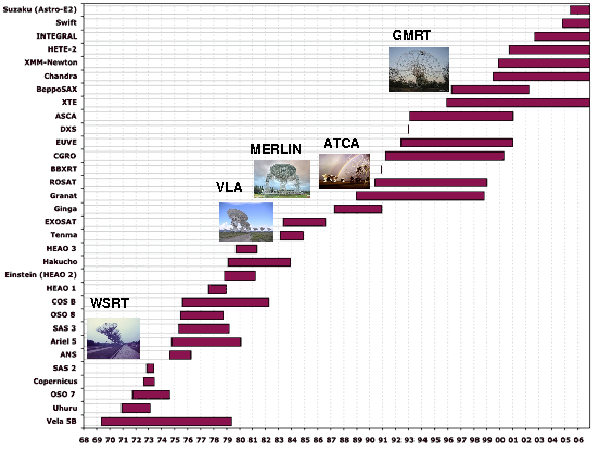}
\caption{An illustration of the comparitive histories of new
  facilities in radio and X-ray astronomy since the 1970s. X-ray
  astronomy has had many more missions, most of which have limited
  lifetimes. Radio astronomy, on the other hand, is still largely
  using facilities and technology from the 1980s. I have not included
  in this figure VLBI arrays; note also that none of the indicated
  radio facilities have been decommissioned -- are all still operating
  as open facilities today. The history of X-ray astronomy is from
  HEASARC ({\bf http://heasarc.gsfc.nasa.gov/}).}
\end{figure}

However, things are about to change. A worldwide renaissance in radio
astronomy is planned, driven in large part by the (almost) united will
of the global radio astronomy community to build the Square Kilometre
Array (SKA, {\bf www.skatelescope.org}), a `transformational'
wide-band world radio facility hopefully to be completed in the 2020s
(see Hall 2005 and Carilli \& Rawlings 2004 for engineering and
science perspectives respectively).

\section{Timeline for the radio renaissance}

The first stage in this radio renaissance will come in the form of
massive increases in the bandwidth of the VLA and MERLIN, resulting in
order of magnitude increases in sensitivity, and the renaming of the
facilities as EVLA and e-MERLIN (`e' is a generic term for upgrade /
expansion / enhancement...). Both facilities are expected to be
operational by 2010. EVLA will have slightly better instantaneous
sensitivity and snapshot imaging, while e-MERLIN will have higher
angular resolution at a given frequency. In addition, the Allen
Telescope Array (ATA), is offering some open time for observations
with its network of small, wide-field, dishes. It is also worth noting
that we have already begun the era of real-time VLBI via the e-EVN and
long baseline array in Australia.

In the overlapping period 2009--2011 three new facilities, which are
rather unlike anything which came before, will begin to explore the
low-frequency radio universe. LOFAR, the largest of these facilities,
will operate in the 30--240 MHz range on baselines up to $\sim 1000$
km. The Long-Wavelength Array (LWA) in the USA will operate similarly
to the LOFAR low-band (30-80 MHz), whereas the Murchison Widefield
Array (MWA) in Australia will operate in the LOFAR high band (80-240
MHz), but without the long baselines. LOFAR in particular is an
extremely flexible instrument which is committed to a large fraction
of open time, and will be discussed in more detail below.

On only a slightly longer timescale, the `1\% SKA' pathfinders,
MeerKAT in South Africa and the Australian SKA Pathfinder (ASKAP),
will be constructed. These telescopes are likely to operate in the GHz
range with conventional, albeit small, dishes, possibly with aperture
arrays. Around the time that their construction is finished, the final
site selection for the SKA is planned to take place, followed by a
`10\% SKA' pathfinder around 2015 and the final SKA by the early
2020s. 

Over this period radio astronomy will have been revolutionized
compared to the position it is now in. Raw sensitivity, fields of view
and observing frequency range will all have increased
dramatically. From the viewpoint of high-energy astrophysics this
means daily monitoring of {\em all known active systems} will be
trivial, and detection of new phenomena via all-sky monitoring at a
wide range fo frequencies will be commonplace. It is useful to think
of the many discoveries brought to high-energy astrophysics by,
e.g. the All-Sky Monitor (ASM) onboard the RXTE satellite. However,
the RXTE ASM can typically detect $\sim 500$ sources per day; on a
similar timescale LOFAR will monitor thousands of sources, and the
final SKA and its pathfinders will be able to track the flux density
variations of {\em millions} of sources every 24 hours.

\begin{figure}
\includegraphics{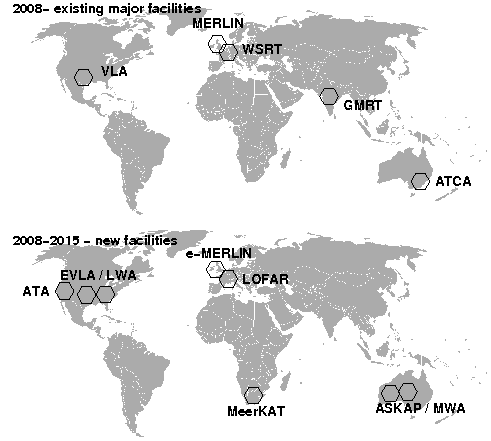}
\caption{An illustration of the new radio facilities planned (and
  funded) within the next decade. The upper map shows the current
  major world astronomical facilities (I have deliberately not
  included VLBI networks, nor ALMA; apologies for other
  omissions). The lower map illustrates all those facilities
  anticipated prior to 2015.  Beyond this timescale, the 10\% and
  finally 100\% SKA will be built in either Australia (ASKAP site) or
  South Africa (MeerKAT site).}
\end{figure}

\section{LOFAR and high-energy astrophysics}

One of the most exciting of the next-generation radio facilities for
high-energy astrophysics, certainly in the next decade (and beyond),
is LOFAR ({\bf www.lofar.org}). LOFAR is a next-generation radio
telescope under construction in The Netherlands with long-baseline
stations under development in other European countries (currently
Germany, The UK, France, Sweden). The array will operate in the 30--80
and 120--240 MHz bands (80--120 MHz being dominated by FM radio
transmissions in northern Europe). The telescope is the flagship
project for ASTRON, and is the largest of the pathfinders for the
lowest-frequency component of the Square Kilometere Array (SKA). Core
Station One (CS1; see Gunst et al. 2006) is currently operating, and
the next stage of deployment is about to begin, with $\sim 50$
stations to be in the field by the end of 2010.

Most of the high-energy astrophysics research areas for LOFAR are
associated with the {\em Transients} Key Science Project (KSP), one of
six KSPs for the initial operations of LOFAR (the others being {\em
  The Epoch of Reionisation}, {\em Surveys}, {\em Cosmic rays /
  particle astrophysics}, {\em Solar / space weather} and {\em Cosmic
  Magnetism}). LOFAR is an exceptional instrument for such astrophysics
because its enormous field of view coupled with multi-beam capability
allows for the first time all-sky monitoring in the radio band, and
because at such low frequencies it efficiently probes coherent radio
emission processes very well for the first time. For more details see
Fender et al. (2008).

\subsection{A census of particle acceleration and kinetic feedback}

\begin{figure}
\includegraphics[width=14cm]{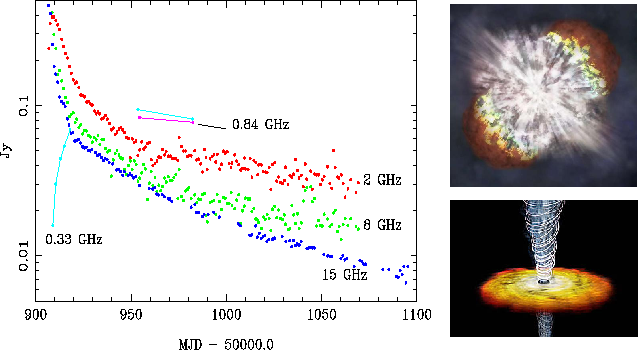}
\caption{Evolution of a radio outburst from the X-ray transient CI
  Cam, with data from the Ryle Telescope (15 GHz), the Green Bank
  Interferometer (2 and 8 GHz) and the Westerbork Synthesis Radio
  Telescope (0.8 and 0.3 GHz).  The radio signal is due to synchrotron
  emission from an expanding source of accelerated electrons. By the
  time of the first radio observations (`externally triggered' by
  X-ray observations) the radio source is already optically thin at
  the highest frequency, 15 GHz. As the source expands, the flux
  density at different frequencies increases as the optical depth at
  that frequency decreases, but then subsequently decays once in the
  optically thin regime as expansion results in energy losses. The
  emission at 330 MHz, just above the LOFAR band, was detectable
  within a few days of the outburst, but did not peak until 20--30
  days later. Such behaviour will be characteristic of explosive
  outburst events associated with e.g. relativistic jets, supernovae,
  GRB afterglows (as indicated by the cartoons to the right of the
  figure), with the rise and decay timescales being an increasing
  function of the luminosity of the event.}
\label{cicam}
\end{figure}

Sychrotron radiation is a key signature of particle acceleration,
which is in turn associated with the explosive injection of energy
into magnetised media on all scales. Common examples include jets from
X-ray binaries, Cataclysmic variables, Active Galactic Nuclei, and
Young Stellar Objects, as well as Supernovae, Gamma-ray bursts and
giant outbursts of Soft Gamma-Ray Repeaters. Fig \ref{cicam} is a
typical example of such an event, in this case for the outburst of a
transient binary system which may contain a black hole.

LOFAR has both advantages and disadvantages when it comes to observing
such radio emission. On the plus side, its enormous field of view and
the slow decay rate of synchrotron plasmas at the low LOFAR
frequencies means that it will really be able to track the 3D
distribution of particle acceleration in the local universe. On the
negative side, as illustrated by Fig \ref{cicam}, it is not the ideal
instrument for rapid triggering of events, as the synchrotron emission
is typically initially self-absorbed at low frequencies and peaks in
flux density much later than the originating explosive event (up to
weeks in X-ray binaries or years in the case of supernovae or GRBs).

\subsection{Pulsars surveys}

\begin{figure}
\includegraphics[width=14cm]{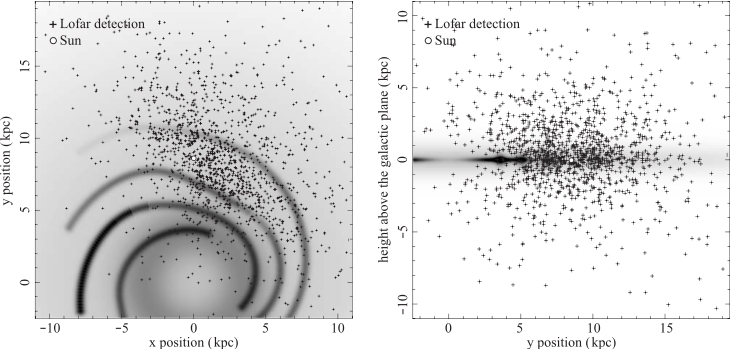}
\caption{The 1000+ pulsars discovered in a 60-day LOFAR all-sky
        survey simulation.  ISM shown in gray. Left) projected on the
        Galactic plane. Right) projected on the plane through the
        Galactic centre and sun, perpendicular to the disk.  From van
        Leeuwen \& Stappers (2008).}
\label{pulsars}
\end{figure}

LOFAR will undertake a major survey of classical radio pulsars as well
as the study of related objects such as Anomalous X-ray Pulsars (AXPs)
and Rotating Radio Transients (RRATs). The LOFAR pulsar survey is
expected to discover more than 1000 new pulsars (see Fig
\ref{pulsars}), which will provide the majority of pulsars for the
    {\em Pulsar Timing Array} (Foster \& Backer 1990) in the northern
    hemisphere. Such a survey also has a fair chance of turning up the
    first pulsar -- black hole binary. In addition, LOFAR will provide
    the sensitivity to allow us to study the individual pulses from an
    unprecedented number of pulsars including millisecond pulsars and
    the bandwidth and frequency agility to study them over a wide
    range which will provide vital new input for models of pulsar
    emission. This will provide us with an unparalleled survey of the
    population of massive star end-products within our galaxy.

\subsection{Extragalactic radio bursts}

LOFAR may detect {\em extragalactic radio bursts}, such as that
reported by Lorimer et al. (2007; Fig 5), to very large distances,
possibly as far as $z \sim 7$, providing a unique probe of the
properties of the intergalactic medium (via their dispersion, and
possibly rotation, measure). Such events may be even be associated
with neutron star--neutron star mergers, in which case the radio
identification of events detectable by advanced LIGO may be
possible. Since such mergers are predicted to provide independent
distance measurements based on their gravitational wave signatures,
identification of the electromagnetic counterpart would provide a
unique test of gravity on cosmological scales, as well as an
independent test of the redshift-distance relation.

\begin{figure}
\includegraphics[width=14cm]{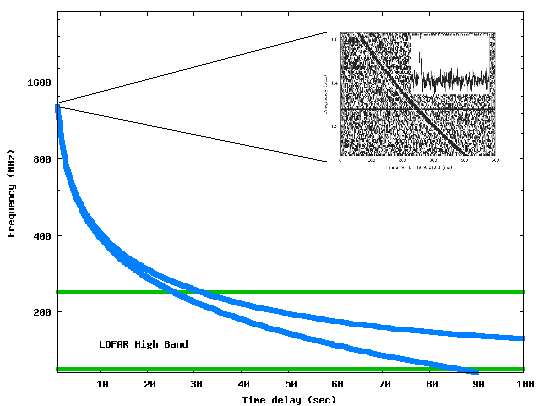}
\caption{An illustration of how the highly-dispersed extragalactic
  radio burst reported by Lorimer et al. (2007) would sweep through
  the LOFAR high band (the frequency limits of which are indicated by
  the green horizontal lines) some tens of seconds later. The blue
  lines indicate the estimated delay and width of the pulse. The
  dispersion delay is assumed to be quadratic; the scatter-broadening
  of the signal is assumed to grow as $\nu^{-4}$ and we assume a pulse
  width of 5 ms at 1.4 GHz (note that the pulse was not resolved by
  Lorimer et al. and so this may be considered to be an {\em upper
  limit} to the scattered pulse width). If the measured steep spectrum
  ($S_{\nu} \propto \nu^{-4}$, the same as the scatter broadening)
  extends to low frequencies, such a burst would be detectable in the
  LOFAR standard data products up to distances in excess of a Gpc,
  allowing unprecedented studies of dispersion and scattering in the
  IGM. Such an event may even have been associated with a neutron
  star--neutron star merger. If so, such events may be detectable by
  LIGO, and a distance inferred from the gravitational wave signal
  alone. LOFAR identification of the host galaxy, via precise
  localisation of the burst, would allow two independent measurements
  of distance on cosmological scales, providing a unique test of
  gravity and of the distance--redshift relatio}
\label{dm3}
\end{figure}

\subsection{Ultra high-energy neutrinos}

LOFAR has already demonstrated its ability to detect radio bursts from
cosmic ray airshowers, which is the main focus of the Cosmic Rays
KSP. However, an even more exciting prospect (also part of the remit
of that KSP) is the detection of radio Cerenkov bursts resulting from
the interaction of high-energy neutrinos with the moon, as predicted
initially by Askaryan (1962). LOFAR is potentially the most sensitive
instrument for this experiment (Scholten et al. 2006; see Connolly
2008 for a summary of other approaches), which could measure
neutrino-nucleon interactions at a centre-of-mass energy exceeding the
Large Hadron Collider by two orders of magnitude (see Fig. 6). Since
the cosmogenic flux of such neutrinos (they are products of the
Greisen-Zatsepin-Kuzmin process which causes the `GZK cutoff' in
cosmic ray energies) is predicted to be very low, such a detection
would either imply `new physics' such as topological defects, or a
nearby strong source of high-energy neutrinos, both of which would be
very exciting results.

\begin{figure}
\includegraphics[width=14cm]{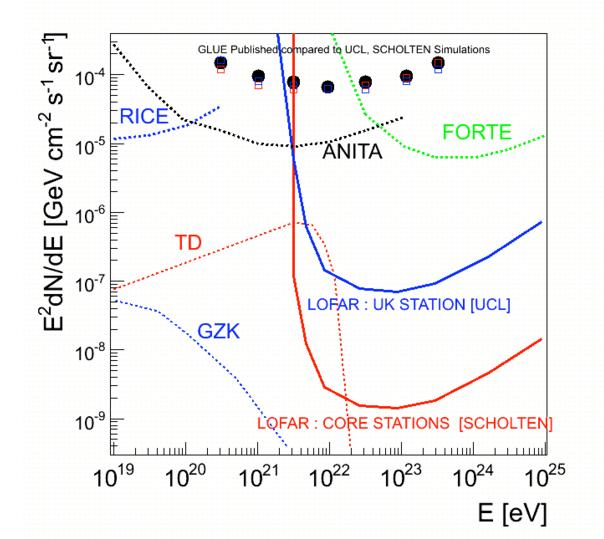}
\caption{Published neutrino limits from RICE, ANITA, FORTE and GLUE
  compared to theoretical expectations from the GZK flux and a
  speculative `new' physics source, Topological Defects (TD). LOFAR
  limits from the UCL simulation of a single UK station and the
  Scholten simulation of the core Dutch array are shown as the solid
  blue and red lines respectively. The predictions (UCL: open blue
  squares; Scholten: open red squares) of the two simulations are also
  compared to the published GLUE limits (closed black
  circles). Courtesy of Mark Lancaster, UCL.}
\label{neu}
\end{figure}

\section{Summary}

High energy astrophysics has benefited greatly on multiple occasions
by combining radio and X-ray data. However, this process has taken
place against a backdrop of ever improving X-ray facilities while the
radio facilities have remained more or less constant in capabilities
for more than two decades. All of this is about to change, with a
whole host of new facilities coming on line in the near future which
will not only deliver improved sensitivity, but revolutionize the way
we do radio astronomy, for example allowing all-sky monitoring for
transient and variable phenomena.

\smallskip

\end{document}